# Predicting inhibitors for SARS-CoV-2 RNA-dependent RNA polymerase using machine learning and virtual screening.

Romeo Cozac[1,*,†], Nazim Medzhidov[1,*,†], and Shinya Yuki[1]

[1]Elix, Inc., Tokyo, Japan
*These authors contributed equally to this work
†Corresponding author. Email: romeo.cozac@elix-inc.com (R.C); nazim.medzhidov@elix-inc.com (N.M)

## Abstract

Global coronavirus disease pandemic (COVID-19) caused by newly identified SARS-CoV-2 coronavirus continues to claim the lives of thousands of people worldwide. The unavailability of specific medications to treat COVID-19 has led to drug repositioning efforts using various approaches, including computational analyses. Such analyses mostly rely on molecular docking and require the 3D structure of the target protein to be available. In this study, we utilized a set of machine learning algorithms and trained them on a dataset of RNA-dependent RNA polymerase (RdRp) inhibitors to run inference analyses on antiviral and anti-inflammatory drugs solely based on the ligand information. We also performed virtual screening analysis of the drug candidates predicted by machine learning models and docked them against the active site of SARS-CoV-2 RdRp, a key component of the virus replication machinery. Based on the ligand information of RdRp inhibitors, the machine learning models were able to identify candidates such as remdesivir and baloxavir marboxil, molecules with documented activity against RdRp of the novel coronavirus. Among the other identified drug candidates were beclabuvir, a non-nucleoside inhibitor of the hepatitis C virus (HCV) RdRp enzyme, and HCV protease inhibitors paritaprevir and faldaprevir. Further analysis of these candidates using molecular docking against the SARS-CoV-2 RdRp revealed low binding energies against the enzyme active site. Our approach also identified anti-inflammatory drugs lupeol, lifitegrast, antrafenine, betulinic acid, and ursolic acid to have potential activity against SARS-CoV-2 RdRp. We propose that the results of this study are considered for further validation as potential therapeutic options against COVID-19.

## Introduction

The global spread of the severe acute respiratory syndrome coronavirus 2 (SARS-CoV-2) has caused the COVID-19 disease pandemic. SARS-CoV-2 has a positive-sense single-strand RNA genome [(+)ssRNA] and belongs to the genus *Betacoronavirus* of the *Coronaviridae* family within the *Nidovirales* order of viruses [1]. The COVID-19 disease is a potentially fatal respiratory disease characterized by atypical pneumonia [2].

SARS-CoV-2 utilizes the receptor-binding domain on the viral spike (S) protein to bind to the angiotensin converting enzyme 2 (ACE2) on host cells and enter the cells via the endosomal pathway [3]. Once inside the cells, viral genomic RNA is released, serving as a template for translation of the viral proteins and for making copies of the viral genome. RNA dependent RNA polymerase (RdRp), also known as non-structural protein 12 (nsp12), is a crucial component of the viral replicase complex responsible for the production of genomic RNA for new virions [4].

Due to the central role of RdRps in the replication of RNA viruses, almost all RNA viruses have an RdRp encoded in their genome. Despite relatively poor sequence similarity, RdRps from different RNA viruses share structural similarity resembling a right hand with thumb, palm, and fingers subdomains, which is also similar to reverse transcriptases. Sequence analyses of RdRps from various viruses revealed conservation of key residues in the active sites as well as in the palm domains [5,6,7,8]. The crystal structure of SARS-CoV-2 RdRp was resolved and recently reported. The general architecture of RdRps, presence of conserved motifs A-G, conservation of key catalytic amino acids as well as structural similarity to Hepatitis C virus (HCV) and poliovirus RdRp was confirmed [9]. RdRps are considered important therapeutic targets due to their crucial role in the viral replication cycle and absence of a counterpart in humans, which can reduce the risk of having undesired side effects during treatment.

A continuous global increase in the COVID-19 cases and unavailability of specific drugs against SARS-CoV-2 urged scientists around the globe to focus on currently available drugs with safety records. FDA urgently approved the use of Remdesivir, an RdRp inhibitor originally developed against the Ebola virus, to treat COVID-19 patients (https://www.fda.gov/news-events/press-announcements/coronavirus-covid-19-update-fda-issues-emergency-use-authorization-potential-covid-19-treatment). However, RNA viruses are known to have high rates of mutations. A recent study identified mutations in the SARS-CoV-2 RdRp gene raising concerns about the potential development of resistance against developed drugs [10]. Thus it is important to consider the possibility that developed treatments might become less effective with time and identify several effective drugs against SARS-CoV-2.

Recent technological developments in the applications of machine learning to drug discovery have shown that it is potentially possible to facilitate the conventional process and reduce the cost for the discovery of new drugs [11,12]. In this study, we implemented machine learning algorithms to identify potential RdRp inhibitors. We trained our models on pre-clinical RdRp inhibitors with inhibitory activity against hepatitis C virus (HCV), poliovirus, dengue virus, and influenza virus RdRps to learn the chemical features of effective RdRp inhibitors from the ligand structures. We then evaluated the ability of our models in identifying known pre-clinical and clinical RdRp inhibitors. Finally, we used our models to screen approved and clinical antiviral and anti-inflammatory drugs to identify potential candidates with inhibitory activity against RdRps. The recently resolved 3D structure of novel coronavirus RdRp [9] provided us with the opportunity to perform molecular docking analysis of the predicted drug candidates against the SARS-CoV-2 RdRp protein active site. Several molecules predicted by machine learning models also showed low binding energies against the target RdRp when analyzed using molecular docking. Some of the predicted molecules are currently in the clinical trials against the novel coronavirus. Our results suggest that in addition to conventional molecular docking approaches, machine learning methods can also be beneficial for drug repositioning purposes and help to narrow down the spectrum of potential drug candidates especially in cases where 3D structure information of the target is not available.

# Methods

## *Data collection*

RdRp inhibitors from PubChem [13] and ChEMBL [14] bioassays were collected in SMILES (Simplified Molecular Input Line Entry System) file format. Collected inhibitors targeted HCV, poliovirus, dengue virus, and influenza virus. Entries with known experimental activity values (IC50/EC50) were selected and assigned binary activity labels based on the activity values to train classification models. The cutoff threshold of activity for training was set at 5μM. The final dataset included 1356 (656 inactive, 700 active) compounds with activity labels. An equal number of active and inactive compounds, amounting to 20**%** of the whole dataset, were randomly selected and used as a validation set. The remaining 80**%** was used as a training set. Unusual compounds which contained only a single atom, or no carbon atoms were removed. Additionally, a test set containing 20 known pre-clinical RdRp inhibitors and 20 presumably unrelated molecules (mostly kinase inhibitors) was established. This set was used to evaluate the general performance of the trained predictive models. Finally, we collected FDA approved and clinical antiviral and anti-inflammatory drugs as test sets for drug repositioning effort.

## *Model Training*

Models utilized in this study can be grouped into 3 categories: classic models, graph-based models, and ensemble models**.** Classic approaches include lasso classification, ridge classification, support vector machines, random forest, shallow neural networks, and xgboost. For these models, compounds in SMILES format were converted to molecular fingerprints using RdKit [15], which in turn were used as input features. We experimented both with circular (Morgan fingerprints) [16] and topological fingerprints. The topological fingerprints were computed by extracting all subgraphs of a compound with a minimum of 1 and a maximum of 7 bonds. To implement the models, we mainly relied on the scikit-learn [17] library. For graph-based models, we tried classic Graph Convolutional Networks, Weave models, and Message Passing Neural Networks (MPNN). We used the DeepChem library to featurize the data and to train the models. The hyperparameters used for training are described in Appendix A1. For MPNN and Weave models, default options worked the best. We have noticed that some of the models have low inter-correlation and are performing well in different circumstances. As a result, using the trained classic and graph based-models, we trained several ensemble models. That is, we combined the output of our base models and trained a new model on top of them, aiming to achieve an overall better performance.

## *Molecular Docking*

Docking analyses were performed using Autodock Vina (v1.1.2) [18] on the resolved crystal structure of SARS-CoV-2 RdRp (PDB ID: 6m71) with the grid box set to the active site [9]. PDB file was converted to PDBQT file format using Aurodock Tools [19] (version 1.5.6), water molecules, and metal ions were removed from the structure and polar hydrogens were added. 3D structures for molecules from antiviral and anti-inflammatory sets were converted to PDBQT file format using OpenBabel v2.4.1 [20]. We used the energy as a scoring function when it was possible. By default, we used the Merck Molecular Force Field (MMFF94) [21]; for a few compounds that contained boron atoms, we used the Universal Force Field (UFF) [22]. When energy minimization failed, the root-mean-square deviation (RMSD) of atomic positions was used instead.

*Environment*

Model training and molecular docking were performed on an NVIDIA DGX Station with 40 CPU cores and 4 V100 GPUs.

## Results and Discussion

The urgent need for effective therapeutic agents against SARS-CoV-2 has resulted in numerous studies focusing on identifying potential drug candidates. A significant amount of *in silico* drug discovery reports has been published recently that propose various candidates for drug repurposing against different protein targets of SARS-CoV-2. Most of those studies utilized conventional molecular docking analyses for which the information on the 3D structure of the target is necessary. Obtaining a 3D crystal structure for the target proteins is a challenging, expensive, and lengthy process. Several artificial intelligence approaches are recently being explored as an alternative that can help researchers find potential drug candidates in a relatively short period even in cases where the 3D structure information is not available. Structural similarities between RdRps of several viruses, conservation of key amino acids in the active site as well as identification of broad-spectrum anti-RdRp drugs like Remdesivir indicated the potential similarity patterns in the chemical structures of effective RdRp inhibitors. This led us to implement supervised machine learning algorithms for the identification of potential RdRp inhibitors. We established a dataset containing small molecules with the experimentally confirmed activity values against RNA dependent RNA polymerases of Hepatitis C Virus (HCV), Dengue virus, Poliovirus, and Influenza virus. This dataset was used to train several classification models to be able to "learn" the chemical properties of effective RdRp inhibitors. Several models achieved an area under the receiver operating characteristic curve (AUROC) score of over 0.8, namely, the Graph Convolutional Network, the Message Passing Network, the Random Forest classifier (both fingerprint types), the Ridge classifier (with circular fingerprints), the Lasso classifier (with topological fingerprints), the 3 layered Multilayer Perceptron (MLP) (with circular fingerprints), and the XGBoost classifier (with topological fingerprints). One of them, the Random Forest classifier on circular fingerprints even surpassed 0.9 in terms of AUROC. In terms of accuracy, the best result of 84% was also observed with the random forest classifier (Table 1.1).

To validate the results, we used 3 best models (based on AUROC score) to run inference on the test set of known pre-clinical RdRp inhibitors. In addition to the AUROC and the accuracy scores, we also report the percentage of true positives, true negatives, false positives, and false negative cases. The result is described in Table 1.2.

**Table 1.1:** Model performance on the validation set.

| Model | AUROC | ACC | Confidence Interval (alpha=0.05) |
|---|---|---|---|
| GraphConv | 0.898 | 0.825 | [0.780, 0.870] |
| Weave | 0.790 | 0.670 | [0.614, 0.726] |
| MPNN | 0.849 | 0.768 | [0.718, 0.818] |
| RandomForest (Circular) | 0.921 | 0.840 | [0.796, 0.884] |
| SVM (Circular) | 0.794 | 0.787 | [0.738, 0.836] |
| Ridge (Circular) | 0.802 | 0.799 | [0.751, 0.847] |
| Lasso (Circular) | 0.752 | 0.742 | [0.690, 0.794] |
| MLP (2 layers) (Circular) | 0.794 | 0.791 | [0.743, 0.839] |
| MLP (3 layers) (Circular) | 0.831 | 0.829 | [0.784, 0.874] |
| XGBoost (Circular) | 0.773 | 0.765 | [0.715, 0.815] |
| RandomForest (Topological) | 0.825 | 0.818 | [0.772, 0.864] |
| SVM (Topological) | 0.780 | 0.772 | [0.722, 0.822] |
| Ridge (Topological) | 0.741 | 0.738 | [0.686, 0.790] |
| Lasso (Topological) | 0.801 | 0.799 | [0.751, 0.847] |
| MLP (2 layers) (Topological) | 0.758 | 0.753 | [0.702, 0.804] |
| MLP (3 layers) (Topological) | 0.725 | 0.715 | [0.661, 0.769] |
| XGBoost (Topological) | 0.816 | 0.810 | [0.763, 0.857] |

*Abbreviations:* AUROC, area under the receiver operating characteristic curve; ACC, accuracy.

**Table 1.2:** Performance of the best 3 models on the test set.

| Model | AUROC | ACC | Confidence Interval (alpha=0.05) | TP | TN | FP | FN |
|---|---|---|---|---|---|---|---|
| **GraphConv** | 0.700 | 0.700 | [0.558, 0.842] | 0.65 | 0.75 | 0.25 | 0.35 |
| **RandomForest (C)** | 0.725 | 0.725 | [0.587, 0.863] | 0.50 | 0.95 | 0.05 | 0.50 |
| **3-layer MLP (C)** | 0.625 | 0.625 | [0.475, 0.775] | 0.50 | 0.75 | 0.25 | 0.50 |

*Abbreviations:* (C), circular fingerprint; AUROC, area under the receiver operating characteristic curve; ACC, accuracy; TP, true positives; TN, true negatives; FP, false positives; FN, false negatives.

Models like the Random Forest classifier were very good at detecting negative examples (a true negative rate of 95%), however, the number of detected positive cases was also affected and the model was able to detect only half of the active molecules (true positive rate of 50%). Other models, like the Graph Convolutional Models, were able to detect more active molecules (true positive rate of 65%), but the true negative rate dropped to 75%, and more false positives were detected. Furthermore, the correlation between the outputs of different models was not very high suggesting that models learned to differentiate the molecules in different ways. Therefore, we thought that an ensemble model might improve the overall performance.

A plain Support Vector Machine with an "RBF" kernel worked the best in our experiments. The model used the outputs of the 10 best models as input features. We first trained the individual models on the original training set. Then, the validation set was split into 2 equal subsets; one of them was used to train the ensemble model, while the second one was set aside as the validation set. The ensemble model slightly outperformed all individual models on the test set (Table 1.3).

**Table 1.3:** Ensemble model results.

| Dataset | AUROC | ACC | Confidence Interval (alpha=0.05) | TP | TN | FP | FN |
|---|---|---|---|---|---|---|---|
| **Validation** | 0.875 | 0.875 | [0.819, 0.931] | 0.871 | 0.879 | 0.129 | 0.121 |
| **Test** | 0.750 | 0.750 | [0.616, 0.884] | 0.600 | 0.900 | 0.100 | 0.400 |

*Abbreviations:* ROC-AUC, area under the receiver operating characteristic curve; ACC, accuracy; TP, true positives; TN, true negatives; FP, false positives; FN, false negatives.

Models were evaluated based on performance on the validation set and a test set of known pre-clinical RdRp inhibitors. Three best performing separate models and the best ensemble model were chosen for the inference analyses. Since only ligand information was used to train the models and no 3D structure of the target SARS-CoV-2 RdRp was used, we wanted to compare our approach to the conventional molecular docking approach, which is based on the 3D structure information of protein target and ligands. We performed virtual screening of the antiviral and anti-inflammatory datasets against the active site of SARS-CoV-2 RdRp (PDB ID: 6m71) using AutoDock Vina [18]. The results of this comparison are described in Table 1.4 and Table 1.5 for antiviral and anti-inflammatory datasets, respectively.

Machine learning models utilized in this study identified several potential drug candidates from both antiviral and anti-inflammatory datasets. As shown in Table 1.4, from the antiviral dataset models were able to identify Remdesivir, a nucleoside analog confirmed to target SARS-CoV-2 RdRp that was recently approved by the US FDA for treatment of COVID-19 patients. Remdesivir was included in the test set as a positive control. Interestingly, baloxavir marboxil, TMC-310911 (ASC09), and umifenovir (Arbidol) identified by our models are currently being investigated in clinical trials for COVID-19. Clinical trial registration and identification numbers are ChiCTR2000029544 for baloxavir marboxil, NCT04261907 for ASC09, and NCT04350684 for umifenovir. Baloxavir marboxil acts on RdRp of Influenza virus [23], while TMC-310911 is a protease inhibitor developed against HIV-1 [24], and umifenovir is an anti-influenza drug that perturbs virus entry into the cells by targeting hemagglutinin (HA) glycoprotein [25]. In addition to these drug candidates, all four of our best performing models identified beclabuvir and asunaprevir as potential RdRp inhibitors from the antiviral dataset. Beclabuvir is a non-nucleoside inhibitor of HCV RdRp (NS5B) [26]. Similar to SARS-CoV-2, HCV is also a single-stranded enveloped positive-sense RNA virus. The active site of both SARS-CoV-2 and HCV RdRp show a degree of structural similarity and they both share the same conserved amino acids in the catalytic site [9]. Binding energy calculations of beclabuvir (-9.2 kcal/mol) towards SARS-CoV-2 RdRp in our experiments also suggest the inhibitory potential of this candidate. Asunaprevir, another anti-HCV drug, is known to target the protease of HCV. Interestingly, all of our best models identified asunaprevir as a potential anti-RdRp candidate, but the binding energy calculation using AutoDock Vina was -7.5 kcal/mol.

Among other candidates both being predicted by at least two of our best models and having relatively low binding energy towards SARS-CoV-2 RdRp were paritaprevir, faldaprevir, simeprevir, vedroprevir (HCV protease inhibitors), ledipasvir, odalasvir, and velpatasvir (HCV NS5A inhibitors). Our models were trained only on RdRp inhibitors, however, several anti-HCV drugs targeting either protease or NS5A protein of the HCV were classified as potential RdRp inhibitors. Interestingly, those candidate molecules also had good binding energy predictions towards SARS-CoV-2 RdRp based on molecular docking analysis.

Effective antiviral drugs can help reduce the viral load in the patients; however, they do not address virus-induced pneumonia directly. This pneumonia is a result of inflammation in the lungs caused by SARS-CoV-2 [27]. Thus COVID-19 patients who have developed pneumonia might need additional therapeutic intervention to suppress the inflammation in the lungs. The ability of our models to identify RdRp inhibitors from the antiviral set motivated us to run the inference analyses on a set of anti-inflammatory drugs. As in the analysis of the antiviral set, we focused on both RdRp inhibitor signature and binding energy predictions against SARS-CoV-2 RdRp. Analysis of the anti-inflammatory dataset revealed that all of our best models predicted betulinic acid and lupeol, both natural products, to possess anti-RdRp activity. Lifitegrast, antrafenine, ursolic acid, dexamethasone acetate, prednisolone phosphate were other candidates predicted by at least two of our models and are also predicted to bind to the active site of SARS-CoV-2 RdRp with binding energy in the range between -7.5 to -9.5 kcal/mol (Table 1.5). Interestingly, both betulinic acid and ursolic acid are pentacyclic triterpenoids with documented antiviral activity against HIV [28].

## Conclusion

In this study, we trained machine learning models on experimentally tested compounds that target the RdRps of several viruses. Models were able to learn the chemical properties of effective inhibitory molecules and predict small molecules with potential activity towards RdRp. Molecular docking analysis of the predicted drug candidates against the catalytic site of the novel coronavirus RdRp revealed low binding energies suggesting that drugs predicted using machine learning algorithms have the potential to bind and inhibit the SARS-CoV-2 RdRp even though only ligand information was used to train those models. We propose the identified small molecules to be considered for experimental validations of their potential activity against SARS-CoV-2 RNA-dependent RNA polymerase.

**Table 1.4:** Antiviral drugs predicted to act on RdRps along with the binding energy values against SARS-CoV-2 RdRp (PDB ID 6m71) calculated using AutoDock Vina.

| Compound | Predicted by # of models | Binding energy to SARS-CoV-2 RdRp (kcal/mol) |
|---|---|---|
| Beclabuvir | 4 | -9.2 |
| Asunaprevir | 4 | -7.5 |
| Paritaprevir | 3 | -10.5 |
| Faldaprevir | 3 | -9.6 |
| Odalasvir | 3 | -8.8 |
| Simeprevir | 3 | -8.7 |
| Vedroprevir | 3 | -8.6 |
| Velpatasvir | 3 | -8.6 |
| Telaprevir | 3 | -8.3 |
| Dolutegravir | 3 | -8.0 |
| Sofosbuvir | 3 | -6.9 |
| Uprifosbuvir | 3 | -6.8 |
| Entecavir | 3 | -6.6 |
| Lobucavir | 3 | -6.6 |
| Trifluridine | 3 | -6.3 |
| Nevirapine | 3 | -6.1 |
| Ledipasvir | 2 | -9.2 |
| Ruzasvir | 2 | -8.1 |
| Baloxavir marboxil | 2 | -8.0 |
| TMC-310911(ASC09) | 2 | -7.9 |
| Adafosbuvir | 2 | -7.8 |
| Remdesivir | 2 | -7.5 |
| Saquinavir | 2 | -7.2 |
| Abacavir | 2 | -7.1 |
| Maribavir | 2 | -7.1 |
| Elvitegravir | 2 | -6.6 |
| Vidarabine | 2 | -6.5 |
| Efavirenz | 2 | -6.3 |
| Valganciclovir | 2 | -6.2 |
| Valomaciclovir | 2 | -6.2 |
| Sorivudine | 2 | -6.1 |
| Ibacitabine | 2 | -6.1 |
| Idoxuridine | 2 | -5.9 |
| Fialuridine | 2 | -5.9 |
| Didanosine | 2 | -5.8 |
| Umifenovir | 2 | -5.8 |

**Table 1.5:** Anti-inflammatory drugs predicted to act on RdRps along with the binding energy values against SARS-CoV-2 RdRp (PDB ID 6m71) calculated using AutoDock Vina.

| Compound | Predicted by # of models | Binding affinity to SARS-CoV-2 RdRp (kcal/mol) |
| --- | --- | --- |
| Betulinic Acid | 4 | -7.4 |
| Lupeol | 4 | -7.2 |
| Lifitegrast | 3 | -9.5 |
| Antrafenine | 3 | -8.7 |
| Ursolic acid | 3 | -8.0 |
| Floctafenine | 3 | -7.1 |
| Cimicoxib | 3 | -7.0 |
| Acemetacin | 3 | -6.8 |
| Morniflumate | 3 | -6.8 |
| Loteprednol | 3 | -6.8 |
| Polmacoxib | 3 | -6.8 |
| Andrographolide | 3 | -6.7 |
| Dexamethasone acetate | 2 | -7.6 |
| Prednisolone phosphate | 2 | -7.5 |
| Cortisone acetate | 2 | -7.3 |
| Mometasone furoate | 2 | -7.3 |
| Prednicarbate | 2 | -7.1 |
| Deflazacort | 2 | -7.1 |
| Clobetasone | 2 | -6.8 |
| Rimexolone | 2 | -6.8 |
| Robenacoxib | 2 | -6.8 |
| Hydrocortisone probutate | 2 | -6.8 |
| Mometasone | 2 | -6.6 |
| Diflunisal | 2 | -6.5 |
| Lumiracoxib | 2 | -6.5 |
| Etoricoxib | 2 | -6.5 |
| Clobetasol | 2 | -6.5 |
| Apremilast | 2 | -6.5 |
| Bisindolylmaleimide I | 2 | -6.5 |
| Talniflumate | 2 | -6.3 |
| NS-398 | 2 | -6.2 |
| Firocoxib | 2 | -5.6 |
| Dimethyl sulfone | 2 | -3.0 |

# Acknowledgments

We would like to thank all healthcare workers, who risk their lives taking care of COVID-19 patients while being forced to live apart from their families.

# Appendix

**A1:** Hyperparameters used for model training.

| Model | Hyperparameters |
|---|---|
| GraphConv | graph_conv_layers=[256, 256], dense_layer_size=128, dropout=0.2, number_atom_features=78 |
| RandomForest | n_estimators=200 |
| SVM | kernel=rbf |
| Ridge Classification | alpha=0.01, loss=hinge |
| Lasso Classification | alpha=0.01, loss=hinge |
| MLP (2 layers) | hidden_layer_sizes=(100,), alpha=0.01 |
| MLP (3 layers) | hidden_layer_sizes=(500, 100), alpha=0.01 |
| XGBoost | max_depth=5, n_estimators=200 |